# Distribution of magnetic domain pinning fields in GaMnAs ferromagnetic films

Jungtaek Kim, D. Y. Shin, and Sanghoon Lee*

*Physics Department, Korea University, Seoul 136-701, Korea*

X. Liu and J. K. Furdyna

*Physics Department, University of Notre Dame, Notre Dame, Indiana 46556, USA*

Using the angular dependence of the planar Hall effect in GaMnAs ferromagnetic films, we were able to determine the distribution of magnetic domain pinning fields in this material. Interestingly, there is a major difference between the pinning field distribution in as-grown and in annealed films, the former showing a strikingly narrower distribution than the latter. This conspicuous difference can be attributed to the degree of non-uniformity of magnetic anisotropy in both types of films. This finding provides a better understanding of the magnetic domain landscape in GaMnAs that has been the subject of intense debate.

*Electronic address: slee3@korea.ac.kr

The ferromagnetic semiconductor GaMnAs continues to be the subject of intense interest due to the opportunities which it may provide for spintronic applications [1-3]. Many of its physical properties, such as the effect of low temperature annealing on the Curie temperature [4-8], the role of Mn ions at interstitial positions [7, 9, 10], and the dependence of magnetic anisotropy on strain and on carrier density, have been extensively investigated [11-13].

The fundamental question of the uniformity of magnetic phase in GaMnAs is, however, still under debate. For example, the ac susceptibility measurements of Hamaya *et al.* [14] of GaMnAs films have led these authors to suggest a mixed magnetic phase model of GaMnAs. Wang *et al.* [15], on the other hand, who also carried out ac susceptibility studies on GaMnAs films, interpret their data in terms of temperature-dependent changes in magnetic anisotropy by assuming a *single* uniform magnetic phase. It is thus clear that further studies are needed to resolve the fundamental issue of the magnetic domain landscape in GaMnAs.

A major advance in this area has been made by Welp *et al*. [16], who used magneto-optical imaging to map the magnetic domain structure in GaMnAs. This study demonstrated a landscape comprised of multiple domains with different directions of magnetization coexisting during magnetization reversal. The magnetotransport measurements of Shin *et al*. [17] provided further evidence for the existence of stable multidomain structures which accompany the magnetization reversal process in GaMnAs.

In this letter we report a study of the multidomain structure in GaMnAs during magnetization reversal carried out by an alternative approach of changing the *direction* of the applied magnetic field while its magnitude is kept constant. It will be shown that this approach provides direct information on the distribution of pinning fields for magnetic domains with different orientations of their local magnetization.

A GaMnAs film with 6.2 % of Mn and a thickness of 100 nm was used in this study. The details regarding growth, annealing, and sample preparation are described elsewhere [19]. Curie temperatures $T_C$ of 62 and 136 K were estimated for the as-grown and annealed films from the temperature dependences of their respective resistivities (see inset in Fig. 4). Hall measurements revealed the magnetic easy axes to lie in the layer plane in both samples. The magnetic anisotropy within the plane is, however, quite different in the two cases, as revealed by the angular dependence of the planar Hall effect (PHE). The dependence of PHE on the magnetic field orientation was measured as follows. The magnetization of the sample was first saturated by applying a field of 4000 Oe at a given angle. The field was then reduced to the desired strength, and the PHE was measured as the field direction was rotated *without changing field strength*. This was repeated for a series of field strengths. The azimuthal angle was measured counterclockwise from $[1\bar{1}0]$ crystallographic direction (i.e., from the direction of the current).

Representative angular dependences of the planar Hall resistance (PHR) obtained for as-grown and annealed GaMnAs films at 13 K are shown in Fig. 1. The PHR is seen to vary between positive and negative values as the field is rotated, showing a hysteresis between clockwise (CW) and counterclockwise (CCW) rotations. This type of angular dependence of PHR has already been observed in GaMnAs films [19, 20], and can be described by the following expression [21]

$$R_{PHR} = \frac{k}{t} M^2 \sin 2\varphi_M , \qquad (1)$$

where $t$ is the film thickness, $\varphi_M$ is the angle between the direction of the current and the magnetization $M$, and $k$ is a constant related to the anisotropic magnetoresistance. Note that the behavior of PHR is strikingly different in the two samples, showing an abrupt transition between its maximum ($+\left|\frac{k}{t}M^2\right|$) and minimum ($-\left|\frac{k}{t}M^2\right|$) values in the as-grown film, and a smooth broad transition for the annealed film.

In order to investigate the magnetic domain structure associated with the reorientation of magnetization during the rotation of the field, we have focused on the transition region indicated by the shaded area in Fig. 1. This region includes the crossing by the field direction of the [110] crystallographic axis. In this angular range the direction of magnetization changes between the [100] and [010] directions via domain nucleation and propagation, in a process similar to that discussed in Ref. [16]. The value of PHR in this region will therefore reflect the fractions of the sample area occupied by domains oriented along [100] and [010]. Note that magnetization in these directions corresponds to the minimum and maximum values of $R_{\text{PHR}}$, respectively, when the entire sample is in single-domain state [17]. The fractional area $p$ of magnetization along the [010] direction at any point during the transition from [100] to [010] (i.e., in the case of CCW rotation of the field) can be calculated from the simple relation, $p = 1/2(1 - R_{\text{PHR}} / R_{\text{PHR}}^{\max})$, given in Ref. [22]. The observed values of $p$ are plotted in the left inset in Fig. 2.

The direction of magnetization in an in-plane magnetized GaMnAs film is determined by the magnetic free energy, given by [23]

$$E = K_{\text{U}} \sin^2 \varphi_M + (K_{\text{C}}/4)\cos^2 2\varphi_M - MH\cos(\varphi_M - \varphi_H), \tag{2}$$

where $H$ is the external magnetic field, $K_{\text{U}}$ and $K_{\text{C}}$ are uniaxial and cubic anisotropy coefficients, respectively, and $\varphi_M$ and $\varphi_H$ are angles of the magnetization and of the applied magnetic field measured from the $[1\bar{1}0]$ direction. The magnetic energy difference between magnetizations along [100] and [010] directions is then

$$\Delta E = E_{[010]} - E_{[100]} = 2MH\cos\varphi_H \sin\left[(\varphi_M^{[010]} - \varphi_M^{[100]})/2\right], \tag{3}$$

where the orientation of magnetizations is given by the angles $\varphi_M^{[100], [010]}$. Note, however, that these orientations may be slightly mis-aligned with [100] and [010] due to the presence of the external magnetic field and to contributions from uniaxial anisotropy [18].

Since the energy difference $\Delta E$ between the two directions of magnetization varies as $H\cos\varphi_H$ in Eq. (3), it can be continuously swept by changing the direction of the applied field. This provides a very direct handle for investigating the pinning energy [24] distribution of magnetic domains, since only those areas having a domain pinning energy of the same or smaller magnitude than $\Delta E$ will make a transition from [100] to [010] during the scan of $\Delta E$. As the sample breaks up into regions with two different directions of magnetization, the value of PHR will change to reflect the fractional areas corresponding to these two magnetization orientations. From the data shown in the left inset in Fig. 2 ($p$ vs. $\varphi_H$) and the relation given in Eq. (3) ($\Delta E$ vs. $\varphi_H$), one can relate the fraction $p$ to the value of $\Delta E/M$, as shown in the right insets in Fig. 2. Note that the relation between $p$ and $\Delta E/M$ is a measure of the field strength required to reorient the domain magnetization from the [100] to the [010] direction. The derivative of $p$ with respect to $\Delta E/M$ then provides the probability of finding a domain with a pinning field $H_p = \Delta E/M$, which is the threshold external field required to cross the [110] direction. The probability distributions of domain pinning fields obtained in this way are plotted for both samples in Fig. 2. This probability can very nicely fit by a Gaussian distribution function given by

$$f(H_p) = \frac{1}{\sigma\sqrt{2\pi}} \exp\left[-\frac{(H_p - H_{avg})^2}{2\sigma^2}\right] , \qquad (4)$$

where $H_p$ is the pinning field for a domain with a given orientation, $H_{avg}$ is the average pinning field over the entire ensemble of domains in the sample, and $\sigma$ is the standard deviation of the pinning field fluctuation. The value of $H_{avg}$ obtained from the fits is $37.7 \pm 0.43$ Oe for the as-grown and $8.6 \pm 0.30$ Oe for the annealed sample.

It is most interesting that the domain pinning field distribution for the as-grown sample shown in Fig. 2 is confined to a very narrow region, indicating that domain pinning is quite homogeneous over the entire sample. Since the domains comprising the sample have nearly identical pinning fields, they rotate coherently during magnetization reversal. This behavior is consistent with the uniform magnetic phase model adopted by Wang *et al.* for the interpretation of their ac susceptibility data obtained on *as-grown* GaMnAs film.

In contrast to the as-grown sample, domain pinning fields in the annealed sample are distributed over a broad region, as seen in Fig. 2. Note that there even is a finite probability *p* for a region of negative pinning fields, indicating that there exist areas in the film in which magnetization direction can change in the absence of an applied magnetic field. The phenomenon of negative pinning fields is usually observed in magnetic multilayer systems comprised of ferromagnetic and antiferromagnetic layers. In that case a strong coupling between adjacent magnetic layers provides a negative pinning field for the ferromagnetic layer [25]. Our sample, however, consists of a single ferromagnetic GaMnAs film, and it is difficult to identify a mechanism that would be analogous to that discussed in Ref. [25]. The negative pinning field in our case must therefore have an entirely different origin.

To gain some insight into the broad distribution of pinning energies in annealed GaMnAs (including negative values), we must reexamine the distribution of magnetic anisotropy within the sample. In the analysis presented above we assume the entire film to be uniformly dominated by the same type of anisotropy (specifically, cubic). However, magnetic domains in a GaMnAs film can in principle have different magnetic anisotropies, resulting in different magnetization directions at zero magnetic field. It is now well established that magnetic anisotropy in GaMnAs is a sensitive function of the hole concentration [11, 13]. Small fluctuations of this parameter can thus produce regions with different magnetic anisotropy (i.e., different relative strengths of $K_U$ and $K_C$). This in turn can result in local variations of magnetic energy profiles, leading to a broad distribution of magnetic pinning fields, as is observed. The observation of the negative pinning field can also be understood in terms of this picture. Assume, for example, that there are some areas dominated by uniaxial anisotropy along [110], and some dominated by cubic anisotropy along the <100> directions. The magnetization of each area will then have tended to align along the directions of magnetic energy minima determined by the anisotropy at zero field. As the magnetization makes the transition from [100] to [010] directions, its relaxation toward the [110] direction in uniaxial-anisotropy-dominated areas can then be viewed as a rotation of magnetization from [100] to [010] in a fraction of the sample. This will manifest itself as a negative pinning energy in the analysis used above, where the magnetization was assumed to make a transition from [100] to [010].

To support the existence of areas dominated by different magnetic anisotropies in the annealed sample, we performed field scans of PHR with different directions of the applied field. Data presented in Fig. 3 show that the value of PHR changes significantly when the field is reduced toward zero *even before* its direction is reversed. This behavior is very different from what we observed in as-grown GaMnAs film with a strong cubic anisotropy, in which an abrupt change of PHR occurs only *after* the field direction is reversed [16,20]. Furthermore, in field scan experiments on annealed GaMnAs films the PHR value at zero field is neither a maximum nor a minimum, which can only occur when the magnetization points along one of the in-plane <100> directions over the entire sample in the form of single domain.

Note also that PHR returns to the *same* value at zero field *independent of what was the field direction* during the scan. These observations directly suggest that the sample consists of some areas with strong uniaxial anisotropy along [110], and some with strong cubic anisotropy along the <100> directions. The magnetization of the sample at zero field is then given by the combination of magnetic domains with magnetizations along [110] (uniaxial-anisotropy-dominated areas) and along <100> (cubic-anisotropy-dominated areas), as shown in the right-hand inset in Fig. 3.

In the above configuration involving domains with two different orientations, the direction of the average magnetization at zero field will lie between these two orientations, thus resulting in an intermediate value of PHR, exactly as observed in experiment. Furthermore, this distribution of magnetic domains is intrinsic to the film, and will thus result in the same value of PHR at zero field, regardless of the direction of the field applied during the scan. This is confirmed by the entire set of field angle data in the left-hand inset of Fig. 3, which clearly show that PHR changes only between two values (positive for [100] and [110] combination, and negative for [010] and [110] combination), both of the same magnitude. This observation directly indicates that the annealed GaMnAs film consists of areas having different magnetic anisotropies, similar to the mixed magnetic phase model suggested by Hamaya *et al*. [14].

The pinning field distribution caused by magnetic fluctuations was further tested by investigating the process of magnetization reorientation for various field strengths, chosen so as to cover the distribution of pinning fields in the sample. If the pinning field distribution shown in Fig. 2 corresponds to a specific distribution of magnetic regions in the sample, only a fraction of the sample with domain pinning fields which are smaller than the applied external field will be able to respond to the field as it is rotated. This will be reflected in the amplitude of PHR as a function of the angle $\varphi_H$, since in the reorientation process only those areas which can follow the rotation of the field will contribute to changes of the PHR value. The field dependence of the amplitude of PHR normalized by the PHR maximum observed at 4000 Oe is shown Fig. 4. The normalized PHR amplitude remains constant (close to unity) at fields above 40 Oe, indicating that above 40 Oe practically the entire sample follows the rotation of the field. This is consistent with the domain pinning field distribution shown in Fig. 2, where the upper ends of the distributions for both samples lie around 40 Oe. At lower fields, on the other hand, the amplitude of PHR is clearly seen to decrease as the field is reduced, since the areas having pinning fields larger than the strength of the applied field can no longer follow the rotation of the field. Thus the lower the field, the lesser fraction of the sample can respond to the field rotation. As seen in Fig. 4, in the as-grown sample the amplitude of PHR drops rapidly to zero within a very narrow field window, while the decrease of PHR in the annealed sample is much more gradual. This behavior directly reflects the difference of domain pinning field distributions in as-grown and annealed GaMnAs.

Interestingly, the pinning field distribution shown in Fig. 2 for GaMnAs films also provides qualitative insight into the temperature dependence of the resistivity shown in the inset of Fig. 4. It is now well established that the peak in the temperature scan of the resistivity is an indication of the magnetic phase transition [6]. The resistance peak in annealed GaMnAs samples -- although it is shifted to a higher temperature, indicating an increase of the Curie temperature -- is systematically observed as a broad, rounded maximum as compared to the sharp peak seen in as-grown samples [4, 6, 8]. This has always been rather surprising, since annealing leads to an overall improvement of magnetic properties of GaMnAs films, such as an enhancement of $T_C$ and a reduction in the concentration of interstitial Mn ions [4-9]. However, one can see from Fig. 2 that the pinning field distribution in annealed GaMnAs is much broader than in as-grown material. This suggests that annealing leads to an increase of magnetic fluctuations (in agreement with Refs. [7] and [14]); and it is reasonable to expect that the more magnetic fluctuations in the film, the broader will be the temperature range needed to complete the magnetic phase transition of the entire film. The commonly observed difference in the resistivity peaks of as-grown and annealed GaMnAs can therefore be readily related to the difference in their magnetic pinning field distributions.

In conclusion, we have investigated the domain pinning field distribution of ferromagnetic GaMnAs films by PHE measurements as a function of applied field orientation. While a narrow pinning field distribution was obtained for as-grown GaMnAs, annealed GaMnAs film exhibited a strikingly broad range of pinning fields. Such pinning field distribution was interpreted in terms of fluctuations of magnetic anisotropy over the area of the film. All domains comprising as-grown GaMnAs have similar magnetic anisotropy (predominantly cubic), resulting in a uniform behavior of the sample as a whole. In contrast, domains in annealed GaMnAs films have a broad range of pinning fields (including even a non-zero probability of *negative* pinning fields), indicating the existence of areas with various degrees of uniaxial anisotropy. This broad pinning field distribution agrees with the mixed magnetic phase model proposed in Ref. [14]. The present study therefore demonstrates that the magnetic nature of GaMnAs films can be represented either by a single (uniform) phase, or by a mixed phase model, depending on the degree of magnetic anisotropy variation related to spatial magnetic fluctuations within the film.

This work was supported by the Korea Research Foundation Grant KRF-2004-005-C00068; by the Seoul R&DB Program; and by the National Science Foundation Grant DMR06-03762.

**Figure Caption**

Fig. 1. Angular dependences of planar Hall resistance data taken at 13 K for (a) as-grown and (b) annealed GaMnAs samples. The open and solid symbols show data taken with field rotations in CW and CCW directions, respectively. A hysteresis is clearly observed between the CW and CCW rotations in both samples, but with very different characteristics.

Fig. 2. Probability of finding magnetic domains oriented along the [010] direction for a given pinning field. The widths of the distribution show a clear contrast between as-grown (solid circles) and annealed (open circles) samples. The solid lines are Gaussian fits obtained with Eq. (4). The left-hand inset shows fractional area $p$ corresponding to magnetic domains with magnetization along [010] obtained from data in the shaded region of Fig. 1. The right-hand insets are plots of the relation between fractional area $p$ and the pinning field $\Delta E/M$ mapped out from the data shown in the left inset.

Fig. 3. Representative field scans of PHR data obtained on annealed GaMnAs film for field directions $\varphi_H = 105°$ and $\varphi_H = 145°$. Both sets of data show significant changes near zero field, giving *the same* intermediate value at the intercept. The zero-field intercepts obtained for PHR scans at different $\varphi_H$ are given in the left-hand inset, showing that the transition occurs between two PHR values that correspond to a combination of two directions of magnetization, along [110] and [100]. This is illustrated in the right-hand inset, where $M_U$ and $M_C$ indicate magnetizations dominated by uniaxial and cubic anisotropies, respectively.

Fig. 4. (a) Amplitude of PHR obtained from angular dependence data taken at various field strengths. The solid and open circles correspond to as-grown and annealed samples. For as-grown sample the PHR amplitude drops to zero in a very narrow field window, and annealed GaMnAs it decreases gradually with decreasing field. Temperature dependence of resistance measured on the two GaMnAs films is shown in the inset.

Fig. 1

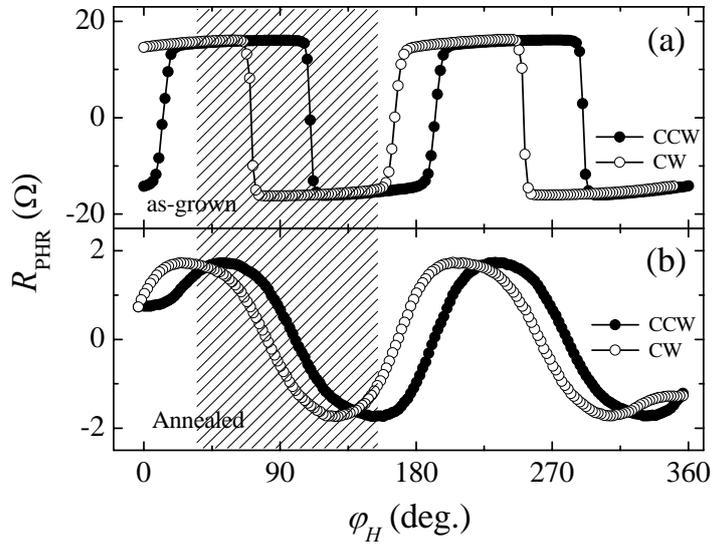

Fig. 2

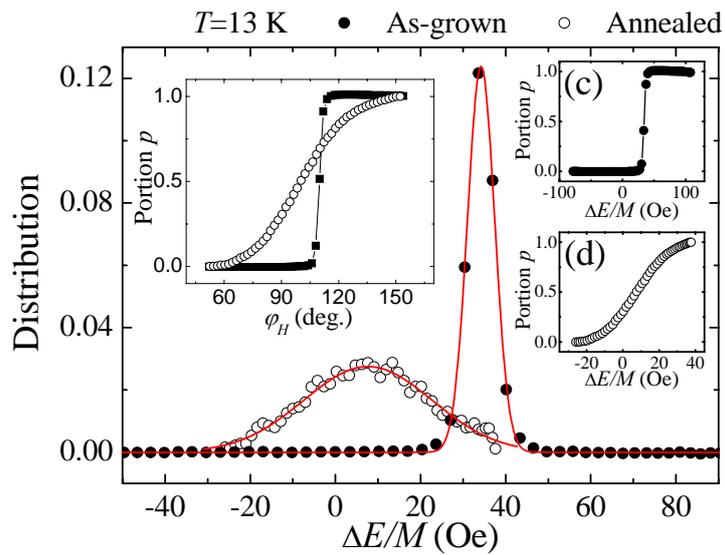

Fig. 3

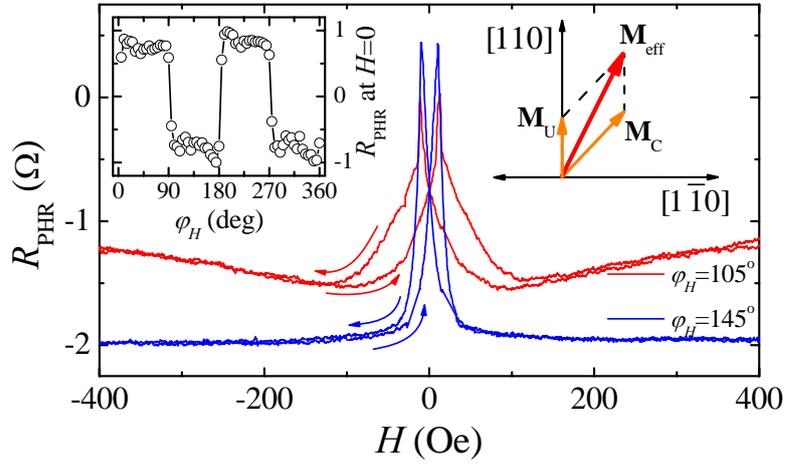

Fig. 4

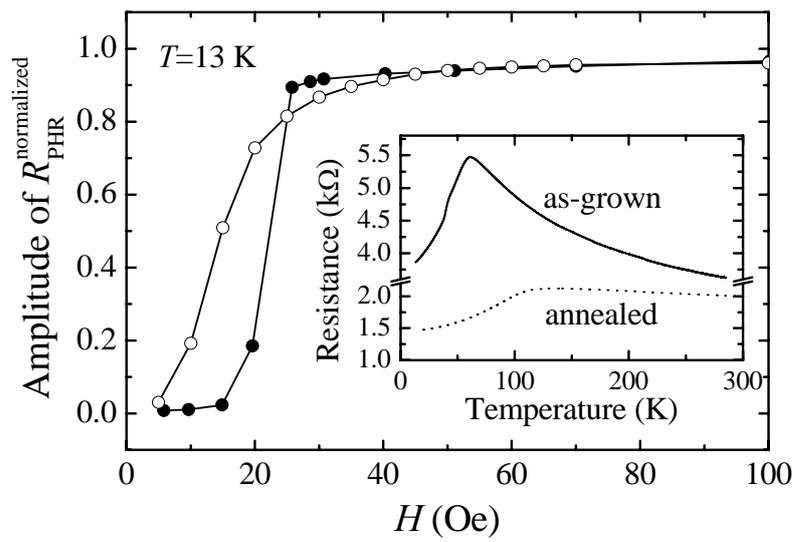